\documentclass[sn-mathphys,Numbered]{sn-jnl}

\usepackage{graphicx}%
\usepackage{multirow}%
\usepackage{amsmath,amssymb,amsfonts}%
\usepackage{amsthm}%
\usepackage{mathrsfs}%
\usepackage[title]{appendix}%
\usepackage{xcolor}%
\usepackage{textcomp}%
\usepackage{manyfoot}%
\usepackage{booktabs}%
\usepackage{algorithm}%
\usepackage{algorithmicx}%
\usepackage{algpseudocode}%
\usepackage{listings}%
\usepackage{enumerate}%
\usepackage{ragged2e}%
\usepackage{indentfirst}%

\theoremstyle{thmstyleone}%
\theoremstyle{thmstyletwo}%

\theoremstyle{thmstylethree}%

\begin{document}

\title[Article Title]{Improvements on ``Multi-Party Quantum Summation without a Third Party based on $d$-Dimensional Bell States''}


\author{\fnm{Xiaobing} \sur{Li}}\email{905035470@qq.com}

\author{\fnm{Jiale} \sur{Hou}}\email{2821541172@qq.com}

\author{\fnm{Haozhen} \sur{Situ}}\email{situhaozhen@gmail.com}

\author*{\fnm{Cai} \sur{Zhang}}\email{zhangcai@scau.edu.cn}

\affil{\orgdiv{College of Mathematics and Informatics}, \orgname{South China Agricultural University}, \orgaddress{\street{No. 483 Wushan Road, Tianhe District}, \city{Guangzhou}, \postcode{510642}, \state{Guangdong}, \country{China}}}


\abstract{In 2021, Wu et al. presented a multi-party quantum summation scheme exploiting the entanglement properties of $d$-dimensional Bell states (Wu et al. in Quantum Inf Process 20:200, 2021). In particular, the authors proposed a three-party quantum summation protocol and then extended their work to a multi-party case. It is claimed that their protocol is secure against outside and participants’ attacks. However, this work points out that Wu's protocol has a loophole, i.e., two or more dishonest participants who meet a specific location relationship can conspire to obtain the private inputs of some honest participants without being detected. Accordingly, improvements are proposed to address these issues.}




\maketitle
     
\section{Introduction}\label{sec1}

Quantum cryptography, combined with classical cryptography and quantum mechanics, has gained much attention since the advent of the first quantum key distribution protocol proposed by Bennett and Brassard \cite{bib22} in 1984, which has thus spawned many branches, such as quantum key distribution (QKD) \cite{bib23,bib24}, quantum secret sharing (QSS) \cite{bib26,bib25}, quantum secure multiparty computation (QSMC) \cite{bib30,bib31,bib27,bib29,bib28}, and so on, and 
QSMC has been developed from the classical secure multiparty computation (MPC). Secure MPC is one of the core technologies to achieve privacy computing in the era of big data, which first originated from the millionaires' problem proposed by Yao \cite{bib1} in 1982, i.e., two millionaires want to know who is richer without exposing their assets. With the quantum algorithms being put forward, such as Shor's algorithm \cite{bib2} and Grover's algorithm \cite{bib9}, and quantum computers' arrival, however, classical MPC that is based on the complexity of the computation is no longer secure, which has led to the research on QSMC.

Quantum secure multiparty summation (QMPS) is one of the subfields in QMPC, whose target is to calculate the summation of inputs from different communicants without the input values leaking out, with the numerous applications in quantum anonymous voting \cite{bib7,bib5}, quantum privacy comparison \cite{bib8}, data mining \cite{bib10}, etc. In 2002, Heinrich \cite{bib3} applied quantum summation into integration. In 2007, Vaccaro et al. \cite{bib11} first applied quantum summation in anonymous voting. To date, various QMPS schemes have been constructed from different viewpoints, such as summation with the help of a third party \cite{bib13}, semi-quantum multi-party summation \cite{bib14}, summation based on single states \cite{bib15} or entangle states \cite{bib32} and so on. In 2019, Ji et al. \cite{bib16} proposed a QMPS protocol based on entanglement swapping with a semi-honest third party. Subsequently, Gan pointed out a loophole in Ji's protocol in Ref \cite{bib17} and presented an improvement on the protocol. Zhang replaced the entanglement swapping between Cat states and Bell states in Ji's protocol with the entanglement swapping between Bell states in Ref \cite{bib18}. Wang et al. \cite{bib19} extended Ji's protocol to sum in decimalization. In addition, QMPS can be classified into tree-type, complete-graph-type, and circle-type, according to the mode of particle transmission \cite{bib12}. In 2021, Wu et al. proposed a quantum summation protocol in the circle-type in Ref \cite{bib21}, which we refer to as Wu's protocol hereafter, and claimed their scheme is secure against the attack pointed out by Liu in Ref \cite{bib20}.

However, it is indicated that Wu's protocol \cite{bib21} has a security loophole in this work, i.e., two or more dishonest participants in their protocol can launch two kinds of attacks to learn about the specific party's private inputs without being detected. To make Wu's protocol secure, we propose improvements by adding random numbers and a detection mechanism.

The remaining part of this paper is organized as follows. Section \ref{sec2} reviews Wu's protocol. Section \ref{sec3} presents two kinds of attacks on Wu's protocol in detail. Section \ref{sec4} puts forward improvements and gives the correctness and security analysis on the improved protocol. Discussion and conclusion are given in the section \ref{sec5}.

\section{Review of Wu's Protocol}\label{sec2}

In Wu's protocol \cite{bib21}, the classical and quantum channels are authenticated, i.e., noiseless and lossless, and the $d$-dimensional Bell state is defined as following : \\
\begin{equation}
\left | \varphi \left ( u,v \right )  \right \rangle = \frac{1}{\sqrt{d} } \sum_{j=0}^{d-1} \omega ^{ju} \left | j  \right \rangle \left | d-j+v  \right \rangle, 
\end{equation}
where $\omega = e^{\frac{2 \pi i }{d} }$, $\ 0 \le u,v \le d-1$ and the operation in Dirac notation is modulo $d$.

The $d$-dimensional Bell states are orthogonal,
\begin{equation}
    \left \langle \varphi (u_{1},v_{1})   | \varphi  (u_{2},v_{2}) \right \rangle  = \delta_{u_{1}u_{2}} \delta_{v_{1}v_{2}}, 
\end{equation}
where
\begin{equation}
\delta_{uv}=\left\{\begin{matrix}
  1,&u=v,  \\
  0,&u\ne v,
\end{matrix}\right.
\end{equation}
is Kronecker delta.

The shifting operation is defined as
\begin{equation}\label{eq.QS}
QS_{k}= \sum_{j=0}^{d-1} \left | j+k  \right \rangle \left \langle j \right|, 
\end{equation}
where the operation in Dirac notation is modulo $d$ $(d\ge2)$ and $ k\in \{0,1,\cdots,d-1\} $.

In Wu’s protocol, $n$ players $A_{1},A_{2},\cdots,A_{n}$ $(n\ge 3)$ are involved. Each participant $A_i$ $(i=1,2,\cdots,n)$ has a private integer string $x_{i}$ of length $m$ in the following form
\begin{equation}\label{eq.x}
    x_{i} = (x_{i1},x_{i2},\cdots,x_{im}),
\end{equation}
where $x_{ij} \in \{0,1,\cdots,d-1\}$ $(j=1,2,\cdots,m;\ d=n-1)$. $A_{1},A_{2},\cdots,A_{n}$ intend to jointly compute the summation $ x_{1} + x_{2} +\cdots+ x_{n}$ (mod $d$) without their private inputs $x_{i}$ leaking out.

The participants agree on the following encoding:\\
\begin{equation}
    v\to \left | \varphi(0,v)   \right \rangle =\frac{1}{\sqrt{d} } \sum_{j=0}^{d-1} \left | j \right \rangle \left | d-j+v \right \rangle \label{eq_2}
\end{equation}
where $v \in \{0,1,2,\cdots,d-1\}$.

The process of the three-party and the multi-party cases in Wu's protocol can be unified as $n$-party $(n\ge 3)$ and described as follows.

\begin{enumerate}[Step 1:]
\item $A_{i}$ $(i=1,2,\cdots,n)$ prepares the initial states as 
\begin{equation*}
    S^{(i)} =\left\{\left | \varphi_{1}^{(i)}    \right \rangle ,\cdots,\left | \varphi_{m}^{(i)}  \right \rangle \right\}
\end{equation*}
according to $x_{i}$, where $\left | \varphi_{j}^{(i)}    \right \rangle =\left | \varphi(0,x_{ij}) \right \rangle$ $(j=1,\cdots,m)$. \\
Subsequently, $A_{i}$ divides the initial states into two sequences
\[ S_{1}^{(i)}  =\left(p_{11}^{(i)},p_{12}^{(i)},\cdots,p_{1m}^{(i)}\right),\]
\[S_{2}^{(i)}  =\left(p_{21}^{(i)},p_{22}^{(i)},\cdots,p_{2m}^{(i)}\right),\]
where $p_{1j}^{(i)}$ and $p_{2j}^{(i)}$ represent the first and the second particles of $j$-th Bell state in $S^{(i)}$ respectively.

\item $A_{i}$ $(i=1,2,\cdots,n)$ mixes $S_{1}^{(i)}$ with $m$ decoy photons $D_{j}^{(i)}$ $(j=1,2,\cdots,m)$ for each particle randomly in $\{\left | r  \right \rangle ,F\left | r \right \rangle \}_{r=0}^{d-1}$ to form new sequences $S_{1}^{'(i)}$, where $F$ is a quantum Fourier transform in $d$-level quantum system. Then $A_{i}$ sends $S_{1}^{'(i)}$ to $A_{i\oplus1}$, and keeps $S_{2}^{(i)}$ in hand, where ``$\oplus$'' satisfies the following arithmetic rules throughout the paper
\begin{align}
    a\oplus b= 
\begin{cases}
  a+b-n, \ &a+b>n, \\
  a+b, \ &a+b\le n,
\end{cases}
\end{align}
and $a,b\in\{1,2,\cdots,n\}$.
\item In this step, $A_{i\oplus1}$ $(i=1,2,\cdots,n)$ utilizes the decoy photons to check whether the transmission of $S_{1}^{'(i)}$ is secure with $A_{i}$. If the error rate is limited in a predeterminded threshold, there is no eavesdropper and the protocol continues; otherwise, the protocol will be terminated.

\item Confirming that there is no eavesdropper in the channel, $A_{i\oplus1}$ $(i=1,2,\cdots,n)$ obtains the sequences $S_{1}^{(i)}$ by discarding the decoy photons. Then $A_{i\oplus1}$ chooses a random sequence $r_{i\oplus1}=\big(r_{(i\oplus1)1},\cdots,r_{(i\oplus1)m}\big) \in \{0,1,\cdots,d-1\}$ and obtains
\begin{align*}
    \bar{S} _{1}^{(i)} &=\left\{QS_{x_{(i\oplus1)1}+r_{(i\oplus1)1}}p_{11}^{(i)},\cdots,QS_{x_{(i\oplus1)m}+r_{(i\oplus1)m}}p_{1m}^{(i)}\right\}
\end{align*}
through performing the shifting operation 
$\big(QS_{x_{(i\oplus1)1}+r_{(i\oplus1)1}},\cdots,QS_{x_{(i\oplus1)m}+r_{(i\oplus1)m}}\big)$ on particle sequence $S_{1}^{(i)}$. Subsequently, $A_{i\oplus1}$ mixes $\bar{S} _{1}^{(i)}$ with $m$ decoy photons $D_{j}^{'(i\oplus1)}$ $(j=1,2,\cdots,m)$ to form new sequence $\ddot{S} _{1}^{(i)}$ and sends it to $A_{i\oplus2}$.

\item In this step, $A_{i\oplus2}$ $(i=1,2,\cdots,n)$ uses the decoy photons to check whether there is an eavesdropper in the channel with $A_{i\oplus1}$. If the error rate is limited in a predeterminded threshold, there is no eavesdropper and the protocol continues; otherwise, the protocol will be terminated.

\item After confirming that there is no eavesdropper in Step 5, $A_{2},\cdots,A_{n},A_{1}$ obtain the particle sequences $\bar{S} _{1}^{(n)},\cdots,\bar{S} _{1}^{(n-2)},\bar{S} _{1}^{(n-1)}$ by discarding the decoy photons. Then $A_{1}$ obtains the classical results $\left(x_{n1}+r_{n1}+s_{11}^{A_{n-1}},\ \cdots,\ x_{nm}+r_{nm}+s_{1m}^{A_{n-1}}\right)$, $\left(s_{21}^{A_{1}},\cdots,s_{2m}^{A_{1}}\right)$ after the $\left | 0  \right \rangle ,\cdots,\left | d-1  \right \rangle$ basis measurement on sequences $\bar{S} _{1}^{(n-1)}$ and $S_{2}^{(1)}$. Subsequently, $A_{1}$ computes and publishes 
\begin{align}
\begin{split}
    P_{1}=&\big(s_{11}^{A_{n-1}}+x_{n1}+r_{n1}+s_{21}^{A_{1}}+(d-x_{11})+(d-r_{11}),\cdots,\\
    &s_{1m}^{A_{n-1}}+x_{nm}+r_{nm}+s_{2m}^{A_{1}}+(d-x_{1m})+(d-r_{1m})\big).
\end{split}
\end{align}
In the same way, $A_{i}$ $(i=2,3,\cdots,n)$ computes and publishes $P_{i}$ respectively as follows:
\begin{align}
    \begin{split}
        P_{2}=&\big(s_{11}^{A_{n}}+x_{11}+r_{11}+s_{21}^{A_{2}}+(d-x_{21})+(d-r_{21}),\cdots,\\
        &s_{1m}^{A_{n}}+x_{1m}+r_{1m}+s_{2m}^{A_{2}}+(d-x_{2m})+(d-r_{2m})\big),\\
        &\cdots,\\
        P_{n}=&\big(s_{11}^{A_{n-2}}+x_{(n-1)1}+r_{(n-1)1}+s_{21}^{A_{n}}+(d-x_{n1})+(d-r_{n1}),\cdots,\\
        &s_{1m}^{A_{n-2}}+x_{(n-1)m}+r_{(n-1)m}+s_{2m}^{A_{n}}+(d-x_{nm})+(d-r_{nm})\big),
    \end{split}
\end{align}
where $s_{1j}^{A_{i}}$ and $s_{2j}^{A_{i}}$ $(i=1,2,\cdots,n;\ j=1,\cdots,m)$ represent the classical results of particles $p_{1j}^{(i)}$ and $p_{2j}^{(i)}$ after $\left | 0  \right \rangle ,\cdots,\left | d-1  \right \rangle$ basis measurement. Here, $s_{1j}^{A_{i}}+s_{2j}^{A_{i}}=x_{ij}$. 
Eventually, each participant obtains the summation of their private integer strings by computing $P_{1}+P_{2}+\cdots+P_{n}$.
\end{enumerate}

\section{Security Analysis of Wu's Protocol}\label{sec3}

Because of the existence of the decoy photons, an outside attack is invalid in Wu's protocol. Conversely, the participant attack should be paid more attention, which is always more powerful referred in Ref \cite{bib33}. In the following, two kinds of participant attacks on Wu's protocol are analyzed in detail. Without loss of generality, we assume that each participant $A_{i}$ $(i=1,2,\cdots,n)$ has $x_{i}$ as input, where $x_{i} \in \{0,1,\cdots,d-1\}$ and $d=n-1$. 
\\ \hspace*{\fill} \\
\noindent \textbf{The Collusive Attack from Two Dishonest Participants} Suppose that $A_{1}$ and $A_{3}$ are the dishonest participants, who work together to obtain $A_{2}$'s private information $x_{2}$. Following the protocol, the computation results $P_{i}$ $(i=1,2,\cdots,n)$ from each participant $A_{i}$ in Step 6 are available as follows.
\begin{align}
\begin{split}
    P_{1}&=s_{1}^{A_{n-1}}+x_{n}+r_{n}+s_{2}^{A_{1}}+(d-x_{1})+(d-r_{1}),\\
P_{2}&=s_{1}^{A_{n}}+x_{1}+r_{1}+s_{2}^{A_{2}}+(d-x_{2})+(d-r_{2}),\\
P_{3}&=s_{1}^{A_{1}}+x_{2}+r_{2}+s_{2}^{A_{3}}+(d-x_{3})+(d-r_{3}),\\
P_{4}&=s_{1}^{A_{2}}+x_{3}+r_{3}+s_{2}^{A_{4}}+(d-x_{4})+(d-r_{4}),\\
&\cdots,\\
P_{n}&=s_{1}^{A_{n-2}}+x_{n-1}+r_{n-1}+s_{2}^{A_{n}}+(d-x_{n})+(d-r_{n}).
\end{split}
\end{align}

Firstly, $A_{1}$ makes the first particle of his own $d$-dimensional Bell state collapse into $s_{1}^{A_{1}}$ by performing the $\{\left | 0  \right \rangle,\left | 1  \right \rangle,\cdots,\left | d-1  \right \rangle \}$ basis measurement in Step 1, and announces $s_{1}^{A_{1}}$ to $A_{3}$. 
In Step 4, $A_{3}$ and $A_{1}$  obtain $s_{1}^{A_{2}}$ and $s_{1}^{A_{n}}$, respectively, 
after the $\{\left | 0  \right \rangle,\left | 1  \right \rangle,\cdots,\left | d-1  \right \rangle \}$ basis measurement on the particles sent from $A_{2}$ and 
$A_{n}$, and then $A_{3}$ tells $s_{1}^{A_{2}}$ to $A_{1}$. Hereafter, 
$A_{3}$ obtains $(s_{1}^{A_{1}}+x_{2}+r_{2})$ after the $\{\left | 0  \right \rangle,\left | 1  \right \rangle,\cdots,\left | d-1  \right \rangle \}$ basis measurement on the particle sent from $A_{2}$ in Step 6. Then, $A_3$ gets the result of $x_{2}+r_{2}$ by computing $\left(s_{1}^{A_{1}}+x_{2}+r_{2}\right)-s_{1}^{A_{1}}$, and announces the result to $A_{1}$. Finally, $A_{1}$ obtains $s_{2}^{A_{2}}$ by calculating $P_{2}-\left(s_{1}^{A_{n}}+x_{1}+r_{1}\right)+(x_{2}+r_{2})$ in Step 6, and he can learn about $A_{2}$'s private input $x_{2}$, where $x_{2}=s_{1}^{A_{2}}+s_{2}^{A_{2}}$. Clearly, $A_{i}$ $(i=1,2,\cdots,n)$ can easily extract $A_{i\oplus1}$'s secret information conspiring with $A_{i\oplus2}$ without being detected when $n\ge4$.
\\ \hspace*{\fill} \\
\noindent \textbf{The Collusive Attack from Four Dishonest Participants} In this kind of attack, no particle is measured to collapse until the measurement in Step 6. Here, we suppose that $A_{1},A_{3},A_{4}$ and $A_{n}$ are the dishonest participants, who work together to launch active attack to acquire $A_{2}$'s secret input $x_{2}$. Firstly, $A_{1},A_{3},A_{4}$ and $A_{n}$ perform the $\{\left | 0  \right \rangle,\left | 1  \right \rangle,\cdots,\left | d-1  \right \rangle \}$ basis measurement on the particles remained in their own hands and record the measurement result respectively in Step 6. After that, $A_{1}(A_{n})$ announces $s_{1}^{A_{1}}(s_{1}^{A_{n}})$ to $A_{3}(A_{1})$ according to their measurement results in the above and their secret information, i.e., $s_{1}^{A_{1}}=x_{1}-s_{2}^{A_{1}}\left(s_{1}^{A_{n}}=x_{n}-s_{2}^{A_{n}}\right)$. Afterward, $A_{3}$ obtains the value of $x_{2}+r_{2}$ by computing $P_{3}-s_{1}^{A_{1}}-s_{2}^{A_{3}}+(x_{3}+r_{3})$, and announces it to $A_{1}$. Meanwhile, $A_{3}$ announces the value of $x_{3}+r_{3}$ to $A_{4}$. $A_{1}$ then obtains the value of $s_{2}^{A_{2}}$ by computing $P_{2}-s_{1}^{A_{n}}-(x_{1}+r_{1})+(x_{2}+r_{2})$ and $A_{4}$ obtains the value of $s_{1}^{A_{2}}$ by computing $P_{4}-(x_{3}+r_{3})-s_{2}^{A_{4}}+(x_{4}+r_{4})$, and announces it to $A_{1}$. Eventually, $A_{1}$ obtains the value of $x_{2}$ by computing $s_{1}^{A_{2}}+s_{2}^{A_{2}}$. It is easy to verify that $A_{i},A_{i\oplus1},A_{i\oplus3}$ and $A_{i\oplus4}$ $(i=1,2,\cdots,n)$ can work together to extract $A_{i\oplus2}$'s secret information without being detected when $n\ge6$.

\section{The Improved Protocols}\label{sec4}

Detailed security analysis in the previous section reveals that Wu's protocol \cite{bib21} has insufficient random numbers for encryption, and lacks detection mechanism to check whether the particles are measured collapsed during the transmission. In this section, the improved protocols are proposed to make Wu's protocol secure and they also remove the secret input encoding operated by each participant in Wu's protocol. The assumptions of the improved protocols are the same as that of Wu's protocol. The detailed revised protocols are as follows.

\subsection{Three-party protocol}\label{sec4_1}

Compared with Wu's protocol, the improved three-party protocol removes the process to encode secret information and random numbers, and reduces the number of times that particles are transmitted in the channel.

\subsubsection{Scheme Description}\label{sec4_1_1}
\begin{enumerate}[Step 1:]
    \item $A_{i}$ $(i=1,2,3)$ prepares the quantum initial state
    \[S^{(i)}=\left\{\left | \varphi_{1}^{(i)} \right \rangle, \left | \varphi_{2}^{(i)} \right \rangle,\cdots,\left | \varphi_{m}^{(i)} \right \rangle \right\},\]
    according to her secret message $x_{i}$ as same as (\ref{eq.x}), where `0' is $\left | \varphi(0,0) \right \rangle$ and `1' is $\left | \varphi(0,1) \right \rangle$. Then $A_{i}$ splits $S^{(i)}$ into two particle sequences
    \begin{align*}
        & S_{1}^{(i)}=\left\{p_{11}^{(i)},p_{12}^{(i)},\cdots,p_{1m}^{(i)}\right\}, \\
        & S_{2}^{(i)}=\left\{p_{21}^{(i)},p_{22}^{(i)},\cdots,p_{2m}^{(i)}\right\},
    \end{align*}
    where $p_{1j}^{(i)}$ and $p_{2j}^{(i)}$ $(j=1,2,\cdots,m)$ represent the first and the second particle of each state in $S^{(i)}$ respectively. After that, $A_{i}$ $(i=1,2,3)$ inserts $m$ decoy photons $D_{j}^{(i)}$, which is randomly in 
    $ \big\{ \left | r \right \rangle, F\left | r \right \rangle\ | \ r \in \{0, 1\} \big\}$, 
    into $S_{1}^{(i)}$ to form a new sequence $S_{1}^{'(i)}$, Then $A_{1}$ ($A_{2}$, $A_{3}$) sends $S_{1}^{'(1)}$ ($S_{1}^{'(2)}$, $S_{1}^{'(3)}$) to $A_{2}$ ($A_{3}$, $A_{1}$) and keeps $S_{2}^{(1)}$ ($S_{2}^{(2)}$, $S_{2}^{(3)}$) in hand.
    \item In this step, $A_{1}$ ($A_{2}$, $A_{3}$) utilizes the decoy photons to check whether the transmission of $S_{1}^{'(1)}$ ($S_{1}^{'(2)}$, $S_{1}^{'(3)}$) is secure with $A_{2}$ ($A_{3}$, $A_{1}$). If the error rate is higher than the predeterminded threshold, they will abort the protocol and restart from Step 1; otherwise, they will proceed to the next step.
    \item $A_{2}$ ($A_{3}$, $A_{1}$) restores $S_{1}^{(1)}$ ($S_{1}^{(2)}$, $S_{1}^{(3)}$) by discarding the decoy photons. Then $A_{1}$ obtains the classical results $\big(s_{11}^{A_{3}},s_{12}^{A_{3}},\cdots,s_{1m}^{A_{3}}\big),\big(s_{21}^{A_{1}},s_{22}^{A_{1}},\cdots,s_{2m}^{A_{1}}\big)$ after performing
    the 
    $\{\left | 0 \right \rangle,\left | 1 \right \rangle \}$ basis measurement on
    $S_{1}^{(3)}$ and $S_{2}^{(1)}$. $A_{1}$ computes and publishes the value of $P_{1}=\Big(s_{11}^{A_{3}}+s_{21}^{A_{1}},s_{12}^{A_{3}}+s_{22}^{A_{1}},\cdots,s_{1m}^{A_{3}}+s_{2m}^{A_{1}}\Big)$. In the same way, $A_{2}$ computes and publishes the results $P_{2}=\Big(s_{11}^{A_{1}}+s_{21}^{A_{2}},s_{12}^{A_{1}}+s_{22}^{A_{2}},\cdots,s_{1m}^{A_{1}}+s_{2m}^{A_{2}}\Big)$. $A_{3}$ computes and publishes the results $P_{3}=\Big(s_{11}^{A_{2}}+s_{21}^{A_{3}},s_{12}^{A_{2}}+s_{22}^{A_{3}},\cdots,s_{1m}^{A_{2}}+s_{2m}^{A_{3}}\Big)$,  where $s_{1j}^{A_{i}}+s_{2j}^{A_{i}}=x_{ij}$ $(i=1,2,3;\ j=1,2,\cdots,m)$.
    \item $A_{1},A_{2},A_{3}$ can obtain the summation of each participant's input by computing $P_{1}+P_{2}+P_{3}$ without their own secret information leaking out.
\end{enumerate}

\subsubsection{Correctness}\label{sec4_1_2}

The target of each participant in the proposed three-party summation protocol is to obtain the value of $x_{1}+x_{2}+x_{3}$ through computing $P_{1}+P_{2}+P_{3}$. In this section, the detail proof is presented. 

$A_{1},A_{2},A_{3}$ agree with the encoding rule in Step 1 beforehand. In Step 3, $A_{1},A_{2},A_{3}$ compute and announce the results as follows:
\begin{align}\label{eq_5}
    \begin{split}
        & P_{1}=\left(s_{11}^{A_{3}}+s_{21}^{A_{1}},s_{12}^{A_{3}}+s_{22}^{A_{1}},\cdots,s_{1m}^{A_{3}}+s_{2m}^{A_{1}}\right), \\
        & P_{2}=\left(s_{11}^{A_{1}}+s_{21}^{A_{2}},s_{12}^{A_{1}}+s_{22}^{A_{2}},\cdots,s_{1m}^{A_{1}}+s_{2m}^{A_{2}}\right), \\
        & P_{3}=\left(s_{11}^{A_{2}}+s_{21}^{A_{3}},s_{12}^{A_{2}}+s_{22}^{A_{3}},\cdots,s_{1m}^{A_{2}}+s_{2m}^{A_{3}}\right), \\
    \end{split}
\end{align}
where $s_{1j}^{A_{i}}$ and $s_{2j}^{A_{i}}$ $(i=1,2,3;\ j=1,2,\cdots,m)$ denote the classical information after the
$\{\left | 0  \right \rangle, \left | 1  \right \rangle \}$ basis measurement on $p_{1j}^{(i)}$ and $p_{2j}^{(i)}$, and $s_{1j}^{A_{i}}+s_{2j}^{A_{i}}=x_{ij}$. As a result, we have
\begin{align*}
    \begin{split}
        P_{1}+P_{2}+P_{3}=&\Big(s_{11}^{A_{3}}+s_{21}^{A_{1}}+s_{11}^{A_{1}}+s_{21}^{A_{2}}+s_{11}^{A_{2}}+s_{21}^{A_{3}},s_{12}^{A_{3}}+s_{22}^{A_{1}}+s_{12}^{A_{1}}+s_{22}^{A_{2}}+s_{12}^{A_{2}}\\
        &+s_{22}^{A_{3}},\cdots,s_{1m}^{A_{3}}+s_{2m}^{A_{1}}+s_{1m}^{A_{1}}+s_{2m}^{A_{2}}+s_{1m}^{A_{2}}+s_{2m}^{A_{3}}\Big) \\
       =&(x_{11}+x_{21}+x_{31},x_{12}+x_{22}+x_{32},\cdots,x_{1m}+x_{2m}+x_{3m}) \\
       =&x_{1}+x_{2}+x_{3},
    \end{split}
\end{align*}
computed by all participants in Step 4 which is the summation of the participants' inputs.

\subsubsection{Security Analysis}\label{sec4_1_3}

The improved protocol is still secure against all kinds of outside attacks since the existence of decoy photons. Consequently, the security against the participant attacks in the protocol should be analyzed in detail. Obviously, any two dishonest participants can gain the honest one's secret input in this three-party protocol. Thus, the case where only one dishonest participant launches attacks is considered.

In the three-party protocol, the role of each party is identical. Without loss of generality, $A_{2}$ is assumed to be the dishonest participant
who attempts to obtain $A_{1}$'s and $A_{3}$'s inputs. To this end, $A_2$ has to learn about the values of $s_{1j}^{A_{1}}+s_{2j}^{A_{1}}$ and $s_{1j}^{A_{3}}+s_{2j}^{A_{3}}$ $(j=1,2,\cdots,m)$. 
$A_{2}$ can get $s_{1j}^{A_{2}}$ and $s_{2j}^{A_{2}}$ by performing the $\{\left | 0  \right \rangle, \left | 1  \right \rangle \}$ basis measurement on $p_{1j}^{(2)}$ and $p_{2j}^{(2)}$, respectively, in Step 1.  In Step 3, $A_{2}$ extracts the values of $s_{1j}^{A_{1}}$ and $s_{2j}^{A_{3}}$ after $A_{1}$ and $A_{3}$ announced $P_{1}$ and $P_{3}$, respectively, according to Eq.(\ref{eq_5}). However, $A_{2}$ can not learn about the exact values of $s_{2j}^{A_{1}}$ and $s_{1j}^{A_{3}}$ from $P_{1}$ published by $A_{1}$. Hence, the participant attack is invalid in the improved protocol.

\subsection{Four or Five-party Protocol}\label{4_2}

In the improved four or five-party protocol, besides removing the process of secret information encoding operated by each participant, we take advantage of detection mechanism to detect dishonest participants. What's more, there is no longer any need for decoy photons in the following improved protocol.

\subsubsection{Scheme Description}\label{sec4_2_1}
\begin{enumerate}[Step 1:]
    \item $A_{i}$ ($i=1,2,\cdots,n$ and $ n=4$ or $n=5$) changes secret information $x_{i}=(x_{i1},x_{i2},\cdots,x_{im})$ into $\overline{x_{i}}=\Big(x_{i1}',x_{i2}',\cdots,x_{i(2m)}'\Big)$, where $x_{i(2j-1)}'+x_{i(2j)}'=x_{ij}$ and $x_{i(2j-1)}',x_{i(2j)}' \in \{0,1,\cdots,d-1\} $ ($j=1,2,\cdots,m;\ d=n-1$). Later, $A_{i}$ encodes $\overline{x_{i}}$ according to Eq.(\ref{eq_2}) and prepares the initial states as
    \[S^{(i)}=\left\{\left | \varphi_{1}^{(i)}   \right \rangle ,\left | \varphi_{2}^{(i)}   \right \rangle ,\cdots,\left | \varphi_{2m}^{(i)}   \right \rangle\right\},\] where $\left | \varphi_{k}^{(i)}  \right \rangle = \left | \varphi(0,x_{ik}')  \right \rangle$ $(k=1,2,\cdots,2m)$. Then, $A_{i}$ divides $S^{(i)}$ into particle sequences
    \[S_{1}^{(i)}=\left\{p_{11}^{(i)},p_{12}^{(i)},\cdots,p_{1(2m)}^{(i)}\right\},\]
    \[S_{2}^{(i)}=\left\{p_{21}^{(i)},p_{22}^{(i)},\cdots,p_{2(2m)}^{(i)}\right\},\] where $p_{1k}^{(i)}$ and $p_{2k}^{(i)}$ indicate the first and the second particle of each state in $S^{(i)}$ respectively, and sends $S_{1}^{(i)}$ to $A_{i\oplus1}$, remaining $S_{2}^{(i)}$ in her hand. 
    \item Upon receiving $S_{1}^{(i)}$ ($i=1,2,\cdots,n$) from $A_{i}$, $A_{i\oplus1}$ selects a group of random sequence $r_{i\oplus1}=\big(r_{(i\oplus1)1},r_{(i\oplus1)2},\cdots,r_{(i\oplus1)2m}\big) \in \left\{0,1,\cdots,d-1 \right\}$. Then, $A_{i\oplus1}$ obtains
    \begin{align*}
       \bar{S}_{1}^{(i)}=\left\{QS_{r_{(i\oplus1)1}}p_{11}^{(i)},QS_{r_{(i\oplus1)2}}p_{12}^{(i)},\cdots,QS_{r_{(i\oplus1)2m}}p_{1(2m)}^{(i)}\right\}
    \end{align*}
    by performing the shifting operation $\big(QS_{r_{(i\oplus1)1}},QS_{r_{(i\oplus1)2}},\cdots,QS_{r_{(i\oplus1)2m}}\big)$ on $S_{1}^{(i)}$ and sends $\bar{S}_{1}^{(i)}$ to $A_{i\oplus2}$.
    \item After receiving $\bar{S}_{1}^{(1)}$ ($\bar{S}_{1}^{(2)}$,$\cdots$, $\bar{S}_{1}^{(n-1)}$, $\bar{S}_{1}^{(n)}$) from $A_{2}$ ($A_{3}$,$\cdots$, $A_{n}$, $A_{1}$), $A_{3}$ ($A_{4},\cdots, A_{1}, A_{2}$) sends the ACK signal to $A_{1}$ ($A_{2},\cdots, A_{n-1}, A_{n}$). Confirming that all participants have received ACK signals, $A_{1}$ selects $t$ particles in $S_{2}^{(1)}$ as checking qudits, denoted as $T^{(1)}$, 
    in which any two particles should not be both $p_{2(2k-1)}^{(1)}$ and $p_{2(2k)}^{(1)}$ $(k=1,2,\cdots,m)$ in $S_{2}^{(1)}$. Note that  
    $T^{(1)}=\left\{p_{2T_{1}^{1}}^{(1)},\cdots,p_{2T_{t}^{1}}^{(1)}\right\}$ and $T_{j}^{1} \in \{1,2,\cdots,2m\}$ ($j=1,2,\cdots,t$).     
    Then, $A_{1}$ sends $T^{(1)}$ to $A_{3}$, and announces the positions of $T^{(1)}$ in $S_{2}^{(1)}$. Thereafter, $A_{2}$ selects $t$ particles in $S_{2}^{(2)}$, denoted as $T^{(2)}$, according to the positions of $T^{(1)}$ in $S_{2}^{(1)}$ published by $A_{1}$ in the following way: (1) if $A_{1}$ takes the $(2k-1)$-th particle in $S_{2}^{(1)}$ as the checking qudit, $A_{2}$ should not take her $2k$-th particle in $S_{2}^{(2)}$ as the checking qudit; (2) if $A_{1}$ takes the $2k$-th particle in $S_{2}^{(1)}$ as the checking qudit, $A_{2}$ should not take her $(2k-1)$-th particle in $S_{2}^{(2)}$ as the checking qudit, where $T^{(2)}=\left\{p_{2T_{1}^{2}}^{(2)},\cdots,p_{2T_{t}^{2}}^{(2)}\right\}$ and $T_{j}^{2} \in \{1,2,\cdots,2m\}$ ($ j=1,2,\cdots,t$). Besides, any two particles in $T^{(2)}$ also should not be both $p_{2(2k-1)}^{(2)}$ and $p_{2(2k)}^{(2)}$ in $S_{2}^{(2)}$. Then $A_{2}$ sends $T^{(2)}$ to $A_{4}$ and publishes the positions of $T^{(2)}$ in $S_{2}^{(2)}$. Subsequently, according to the same way as $A_{1}$ and $A_{2}$, $A_{3},\cdots,A_{n}$ select $t$ particles in $S_{2}^{(3)},\cdots,S_{2}^{(n)}$ in order, and send $T^{(3)},\cdots,T^{(n)}$ with their positions in $S_{2}^{(3)},\cdots,S_{2}^{(n)}$ to $A_{3\oplus2},\cdots,A_{2}$, where $T^{(i)} = \Big\{p_{2T_{1}^{i}}^{(i)},\cdots,p_{2T_{t}^{i}}^{(i)}\Big\}$ $(i=3,\cdots,n)$.
    \item In this Step, $A_{i\oplus1}$ $(i=1,2,\cdots,n)$ needs to divulge the values of $\big(r_{(i\oplus1)T_{1}^{1}},\cdots,r_{(i\oplus1)T_{t}^{1}}\big)$ after $A_{i\oplus2}$ receiving $T^{(i)}$ with its position in $S_{2}^{(i)}$. Then, $A_{i\oplus2}$ performs $d$-dimensional Bell measurement on particle pairs $\Big(p_{1T_{j}^{i}}^{(i)'},p_{2T_{j}^{i}}^{(i)}\Big)$, and announces the measurement results to $A_{i}$, where $p_{1T_{j}^{i}}^{(i)'}=QS_{r_{(i\oplus1)T_{j}^{i}}}p_{1T_{j}^{i}}^{(i)}$ $(j=1,2,\cdots,t)$. $A_{i}$ checks whether the measurement results announced by $A_{i\oplus2}$ are consistent with $\left | \varphi\left(0,x_{iT_{j}^{1}}'+r_{(i\oplus1)T_{j}^{1}} \right) \right \rangle$. If the error rate surpasses the predefined threshold, the protocol will be aborted; otherwise, the protocol continues.
    \item Confirming that there is no malicious participant in Step 4, $A_{i\oplus2}$ $(i=1,2,\cdots,n)$ informs the measurement results to $A_{i}$ after the $\{\left | 0 \right \rangle,\left | 1 \right \rangle,\cdots,\left | d-1 \right \rangle \}$ basis measurement on $T^{(i)}$. Thereafter, $A_{1}$ obtains the sequence of classical results $\Big(s_{11}^{A_{n-1}}+r_{n1},\cdots,s_{1(2m)}^{A_{n-1}}+r_{n(2m)}\Big)$, $\Big(s_{21}^{A_{1}},\cdots,s_{2(2m)}^{A_{1}}\Big)$ through performing the $\{\left | 0 \right \rangle ,\left | 1 \right \rangle ,\cdots,\left | d-1 \right \rangle \}$ basis measurements on $\bar{S}_{1}^{(n-1)}$ and $S_{2}^{(1)}$. Then, $A_{1}$ computes and publishes
    \begin{align}\label{eq_P1}
        \begin{split}
            P_{1}=&\Big(s_{11}^{A_{n-1}}+r_{n1}+s_{21}^{A_{1}}+(d-r_{11}),\cdots,s_{1(2m)}^{A_{n-1}}+r_{n(2m)}+s_{2(2m)}^{A_{1}}\\
            &+(d-r_{1(2m)})\Big).
        \end{split}
    \end{align}
    In the same way, $A_{i}$ $(i=2,\cdots,n)$ compute and publish the following results:
    \begin{align}\label{eq_4}
        \begin{split}
            P_{2}=&\Big(s_{11}^{A_{n}}+r_{11}+s_{21}^{A_{2}}+(d-r_{21}),\cdots,s_{1(2m)}^{A_{n}}+r_{1(2m)}+s_{2(2m)}^{A_{2}}\\
            &+(d-r_{2(2m)})\Big),\\
            &\cdots, \\
            P_{n}=&\Big(s_{11}^{A_{n-2}}+r_{(n-1)1}+s_{21}^{A_{n}}+(d-r_{n1}),\cdots,s_{1(2m)}^{A_{n-2}}+r_{(n-1)2m}\\
            &+s_{2(2m)}^{A_{n}}+(d-r_{n(2m)})\Big).
        \end{split} 
    \end{align}
    \item $A_{1},\cdots,A_{n}$ compute $P_{1}+P_{2}+\cdots+P_{n}=(T_{1},\cdots,T_{2m})$, where $\left(T_{1}+T_{2},T_{3}+T_{4},\cdots,T_{2m-1}+T_{2m}\right)$ is the summation of each participant's secret input.
\end{enumerate}

\subsubsection{Correctness}\label{sec4_2_2}

In Step 6 of the above protocol, $A_{1},A_{2},\cdots,A_{n}$ ($n=4$ or $5$), according to Eq.(\ref{eq_P1},\ref{eq_4}), calculate the value of 
\begin{align}\label{eq_6}
    \begin{split}
        P_{1}+P_{2}+\cdots+P_{n}=&\left ( s_{11}^{A_{n-1}}+r_{n1}+s_{21}^{A_{1}}+(d-r_{11})+s_{11}^{A_{n}}+r_{11}+s_{21}^{A_{2}}+(d-r_{21}) \right. \\
         &\left. +\cdots+s_{11}^{A_{n-2}}+r_{(n-1)1}+s_{21}^{A_{n}}+(d-r_{n1}),\ \cdots\ ,\right. \\
         &\left. s_{1(2m)}^{A_{n-1}}+r_{n(2m)}+s_{2(2m)}^{A_{1}}+(d-r_{1(2m)})+s_{1(2m)}^{A_{n}}+r_{1(2m)}+s_{2(2m)}^{A_{2}} \right. \\
         &\left. +(d-r_{2(2m)})+\cdots+s_{1(2m)}^{A_{n-2}}+r_{(n-1)2m}+s_{2(2m)}^{A_{n}}+(d-r_{n(2m)}) \right ) \\
         =& \left ( x_{11}'+x_{21}'+\cdots+x_{n1}',\cdots,x_{1(2m)}'+x_{2(2m)}'+\cdots+x_{n(2m)}' \right ),
    \end{split}
\end{align}
where $x_{i(2j-1)}'+x_{i(2j)}'=x_{ij}$ ($i=1,2,\cdots,n;\  j=1,2,\cdots,m$) and $\left( (x_{11}'+x_{21}'+\cdots+x_{n1}')+(x_{12}'+x_{22}'+\cdots+x_{n2}'),\cdots,(x_{1(2m-1)}'+x_{2(2m-1)}'+\cdots \right. \\
\left. +x_{n(2m-1)}')+(x_{1(2m)}'+x_{2(2m)}'+\cdots+x_{n(2m)}') \right)$ is the summation of all participants' inputs and the correctness of the improved protocol guaranteed.

\subsubsection{Security Analysis}\label{sec4_2_3}

In this section, it is first shown that the improved protocol is immune to the collusive attack from two participants referred in Sect.\ref{sec3}. Without loss of generality, $A_{1},A_{3}$ are assumed to be the dishonest participants, aiming to obtain $A_{2}$'s secret input. 
To this end, 
$A_{3}$ has to perform the $\{\left | 0  \right \rangle,\left | 1  \right \rangle,\cdots,\left | d-1  \right \rangle \}$ basis measurement on particle $p_{1j}^{(2)}$ to obtain $s_{1j}^{A_{2}}$ ($j=1,2,\cdots,2m$), which will be detected with the probability of $\frac{t}{2m}$ in Step 4. Further, if an outside eavesdropper launches active attacks on the particle transmitted in the channel, he will be also detected with the same probability in Step 4.

In the case of the collusive attack from $n-2$ participants, except for $A_{k}$ and $A_{l}$ ($k,l \in \{1,2,\cdots,n\}$ and $k < l$), the remaining $n-2$ participants are assumed to be attackers. If $l=k+2$ or $l=k+n-2$, according to Eq.(\ref{eq_P1},\ref{eq_4}), the remaining $n-2$ participants can obtain the value of $s_{1j}^{A_{k}}+s_{2j}^{A_{l}}$ in Step 5, but they can not deduce $x_{kj}$ and $x_{lj}$ ($j=1,2,\cdots,2m,$) without knowing the exact values of $s_{1j}^{A_{k}}$ and $s_{2j}^{A_{l}}$. If $l \ne k+2$ and $l \ne k+n-2$, the remaining $n-2$ participants can obtain the values of $s_{1j}^{A_{l}}+r_{kj}$ or $s_{2j}^{A_{l}}+r_{kj}$, but they can not obtain $x_{lj}$  without the exact value of $r_{kj}$. In conclusion, the improved protocol is secure against the participant attacks.

\subsection{Multi-party Protocol}\label{sec4_3}

Based on the revised protocol described in the above section, we expand the four or five-party protocol to multi-party scenarios by slightly changing the steps. In addition to the same improvement as the protocol described above, the primary disparity between Wu's protocol and the improved protocol in multi-party case is the increase in the number of random numbers.

\subsubsection{Scheme Description}\label{sec4_3_1}
\begin{enumerate}[Step 1:]
    \item $A_{i}$ ($i=1,2,\cdots,n$ and $n \ge 6$) splits secret message $x_{i}=(x_{i1},\cdots,x_{im})$ into $\overline{x_{i}}=\big(x_{i1}',x_{i2}',\cdots,x_{i(2m)}'\big)$, where $x_{i(2j-1)}'+x_{i(2j)}'=x_{ij}$ and $x_{i(2j-1)}',x_{i(2j)}'\in\{0,1,\cdots,d-1\}$ $(j=1,2,\cdots,m;\ d=n-1)$. Then, $A_{i}$ encodes $\overline{x_{i}}$ according to Eq.(\ref{eq_2}) and prepares the initial states as \[S^{(i)}=\left\{\left | \varphi_{1}^{(i)}   \right \rangle ,\left | \varphi_{2}^{(i)}   \right \rangle ,\cdots,\left | \varphi_{2m}^{(i)}   \right \rangle \right\},\] where $\left | \varphi_{k}^{(i)}  \right \rangle = \left | \varphi(0,x_{ik}') \right \rangle$ $(k=1,2,\cdots,2m)$. Further, $A_{i}$ divides $S^{(i)}$ into particle sequences \[S_{1}^{(i)}=\left\{p_{11}^{(i)},p_{12}^{(i)},\cdots,p_{1(2m)}^{(i)}\right\},\]
    \[S_{2}^{(i)}=\left\{p_{21}^{(i)},p_{22}^{(i)},\cdots,p_{2(2m)}^{(i)}\right\},\] where $p_{1k}^{(i)}$ and $p_{2k}^{(i)}$ represent the first and the second particle of each Bell state in $S^{(i)}$ respectively. After that, $A_{i}$ prepares $\lambda-1$ $(\lambda=\left \lfloor \frac{n}{2}  \right \rfloor)$ groups of random sequences $r_{i}^{g}=\left(r_{i1}^{g},r_{i2}^{g},\cdots,r_{i(2m)}^{g}\right) \in \{0,1,\cdots,d-1\}$ ($g=1,2,\cdots,\lambda-1$). Then, $A_{i}$ obtains
    \begin{align*}
        S_{1}^{(i)1}=\left\{QS_{(d-r_{i1}^{1})}p_{11}^{(i)}, QS_{(d-r_{i2}^{1})}p_{12}^{(i)},\cdots,QS_{(d-r_{i(2m)}^{1})}p_{1(2m)}^{(i)}\right\}
    \end{align*}
    through performing the shifting operation $\Big(QS_{(d-r_{i1}^{1})},\cdots,QS_{(d-r_{i(2m)}^{1})}\Big)$ on $S_{1}^{(i)}$. Afterward, $A_{i}$ sends $S_{1}^{(i)1}$ to $A_{i\oplus1}$ and keeps $S_{2}^{(i)}$ in hand. 
    \item $A_{i\oplus1}$ $(i=1,2,\cdots,n)$ obtains $S_{1}^{(i)2}$ by performing the shifting operation $\Big(QS_{r_{(i\oplus1)1}^{1}},\cdots,QS_{r_{(i\oplus1)2m}^{1}}\Big)$ on sequence $S_{1}^{(i)1}$ after receiving $S_{1}^{(i)1}$, and sends $S_{1}^{(i)2}$ to $A_{i\oplus2}$. $A_{i\oplus2},\cdots,A_{i\oplus (\lambda-1)}$ do the same as $A_{i\oplus1}$ in turn; namely, $A_{i\oplus2},\cdots,A_{i\oplus (\lambda-1)}$ take turns to perform $\Big(QS_{r_{(i\oplus2)1}^{2}},\cdots,QS_{r_{(i\oplus2)2m}^{2}}\Big),\cdots,\Big(QS_{r_{(i\oplus(\lambda-1))1}^{\lambda-1}},\cdots,QS_{r_{(i\oplus(\lambda-1))2m}^{\lambda-1}}\Big)$ on the particle sequences sent by the previous participants. The specific transmission process is shown in Fig.\ref{Fig_1}. Finally, $A_{i\oplus\lambda}$ receives $S_{1}^{(i)\lambda}$ from $A_{i\oplus(\lambda-1)}$, where
    \begin{align}\label{eq_14}
        \begin{split}
            S_{1}^{(i)\lambda}= \Big\{&QS_{r_{(i\oplus \lambda-1)1}^{\lambda-1}}\cdots QS_{r_{(i\oplus1)1}^{1}}QS_{(d-r_{i1}^{1})}p_{11}^{(i)},\cdots,\\
            &QS_{r_{(i\oplus \lambda-1)2m}^{\lambda-1}}\cdots QS_{r_{(i\oplus1)2m}^{1}}QS_{(d-r_{i(2m)}^{1})}p_{1(2m)}^{(i)}\Big\}, 
        \end{split}
    \end{align}
    for simplicity, denoting Eq.\ref{eq_14} as 
    \begin{align}
        S_{1}^{(i)\lambda}=\{p_{11}^{(i)'},\cdots,p_{1(2m)}^{(i)'}\}.
    \end{align}
    Then $A_{i\oplus \lambda}$ sends ACK signal to $A_{i}$.
    \item After $A_{1},\cdots,A_{n}$ receiving ACK signals from $A_{1\oplus \lambda},\cdots,A_{n\oplus \lambda}$, $A_{1}$ randomly selects $t$ particles in $S_{2}^{(1)}$ as checking qudits, denoted as $T^{(1)}$, in which any two particles should not be both $p_{2(2k-1)}^{(1)}$ and $p_{2(2k)}^{(1)}$ in $S_{2}^{(1)}$, where $T^{(1)}=\left\{p_{2T_{1}^{1}}^{(1)},\cdots,p_{2T_{t}^{1}}^{(1)}\right\}$ and $T_{j}^{1} \in \{1,2,\cdots,2m\}$ ($j=1,2,\cdots,t;\ k=1,2,\cdots,m$). Then, $A_{1}$ sends $T^{(1)}$ to $A_{1\oplus \lambda}$, and announces the positions of $T^{(1)}$ in $S_{2}^{(1)}$ publicly. Thereafter, $A_{2}$ selects $t$ particles in $S_{2}^{(2)}$, denoted as $T^{(2)}$, according to the positions of $T^{(1)}$ in $S_{2}^{(1)}$ published by $A_{1}$ in the following way: (1) if $A_{1}$ takes the $(2k-1)$-th particle in $S_{2}^{(1)}$ as checking qudit, $A_{2},\cdots,A_{\lambda}$ should not take their $2k$-th particles in $S_{2}^{(2)},\cdots,S_{2}^{(\lambda)}$ as the checking qudits; (2) if $A_{1}$ takes the $2k$-th particle in $S_{2}^{(1)}$ as the checking qudit, $A_{2},\cdots,A_{\lambda}$ should not take their $(2k-1)$-th particles in $S_{2}^{(2)},\cdots,S_{2}^{(\lambda)}$ as the checking qudits, where $T^{(2)}=\left\{p_{2T_{1}^{2}}^{(2)},\cdots,p_{2T_{t}^{2}}^{(2)}\right\}$ and $T_{j}^{2} \in \{1,2,\cdots,2m\}$ $(j=1,2,\cdots,t;\ k=1,2,\cdots,m)$. Besides, any two particles in $T^{(2)}$ also shouldn't be both $p_{2(2k-1)}^{(2)}$ and $p_{2(2k)}^{(2)}$. Then $A_{2}$ sends $T^{(2)}$ to $A_{2\oplus\lambda}$ and publishes the positions of $T^{(2)}$ in $S_{2}^{(2)}$. Subsequently, $A_{3},\cdots,A_{n}$ select $t$ particles in $S_{2}^{(3)},\cdots,S_{2}^{(n)}$ 
    \begin{figure}
        \centering
        \includegraphics[scale=0.45]{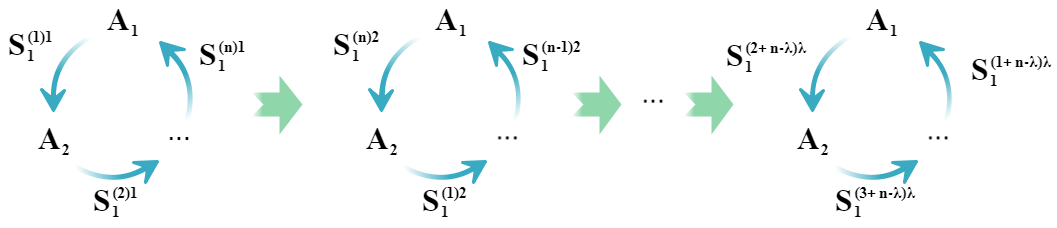}
        \caption{The flowchart of the particle sequences transmission in the improved multi-party protocol. It's worth noting that each sequence is performed shifting operators and transported $\lambda$ times in total.}
        \label{Fig_1}
    \end{figure}
    in order, according to the same rules as $A_{1}$ and $A_{2}$, and send $T^{(3)},\cdots,T^{(n)}$ to $A_{3\oplus \lambda},\cdots,A_{n\oplus \lambda}$ with their positions in $S_{2}^{(3)},...,S_{2}^{(n)}$, where $T^{(i)}=\Big\{p_{2T_{1}^{i}}^{(i)},\cdots,p_{2T_{t}^{i}}^{(i)}\Big\}$ and $T_{j}^{i} \in \{1,2,\cdots,m\}$ $(i=3,\cdots,n;\ j=1,2,\cdots,t)$.
    \item In this Step, $A_{i\oplus1},\cdots,A_{i\oplus (\lambda-1)}$ $(i=1,2,...,n)$ need to publish the values of $\Big(r_{(i\oplus1)T_{1}^{i}}^{1},\cdots,r_{(i\oplus1)T_{t}^{i}}^{1}\Big),\cdots,\Big(r_{(i\oplus \lambda-1)T_{1}^{i}}^{\lambda-1},\cdots,r_{(i\oplus \lambda-1)T_{t}^{i}}^{\lambda-1}\Big)$ respectively, where $T_{j}^{i} \in \{1,2,\cdots,2m\}$ $(j=1,2,\cdots,t)$. Then, $A_{i\oplus \lambda}$ performs $d$-dimensional Bell measurement on particle pairs $\Big(p_{1T_{j}^{i}}^{(i)'}$, $p_{2T_{j}^{i}}^{(i)}\Big)$ and informs the measurement results to $A_{i}$. $A_{i}$ checks whether the measurement results are consistent with $\left | \varphi\left(0,x_{iT_{j}^{i}}'+(d-r_{iT_{j}^{i}}^{1})+r_{(i\oplus1)T_{j}^{i}}^{1}+\cdots+r_{(i\oplus \lambda-1)T_{j}^{i}}^{\lambda-1}\right)  \right \rangle$ $(j=1,2,\cdots,t)$. If the error rate is higher than the predefined threshold, $A_{i}$ will abort the protocol; otherwise, the protocol continues.
    \item Upon confirming that there is no dishonest participant in Step 4, $A_{i\oplus \lambda}$ $(i=1,2,\cdots,n)$ informs the measurement results after $\left | 0 \right \rangle,\left | 1 \right \rangle,\cdots,\left | d-1 \right \rangle $ basis measurement on $T^{(i)}$ to $A_{i}$. Then, $A_{1},\cdots,A_{n}$ perform $\left | 0 \right \rangle ,\left | 1 \right \rangle ,\cdots,\left | d-1 \right \rangle $ basis measurements on $\left(S_{1}^{(1+ n-\lambda)\lambda},S_{2}^{(1)}\right),\cdots,\left(S_{1}^{(n-\lambda)\lambda},S_{2}^{(n)}\right)$ respectively. Eventually, $A_{1},...,A_{n}$ compute and publish the following results:
    \begin{align}\label{eq_16}
        \begin{split}
            P_{1}=\Bigg(&s_{11}^{A_{1+n-\lambda}}+r_{(2+n-\lambda)1}^{1}+\cdots+r_{n1}^{\lambda-1}+s_{21}^{A_{1}}+\Big(d-r_{(1+n-\lambda)1}^{1}\Big)\\
            &+\sum_{g=2}^{\lambda-1}(d-r_{11}^{g}),\cdots,s_{1(2m)}^{A_{1+n-\lambda}}+r_{(2+n-\lambda)2m}^{1}+\cdots+r_{n(2m)}^{\lambda-1}\\
            &+s_{2(2m)}^{A_{1}}+\Big(d-r_{(1+n-\lambda)2m}^{1}\Big)+\sum_{g=2}^{\lambda-1}\Big(d-r_{1(2m)}^{g}\Big)\Bigg),\\
            \\
            P_{2}=\Bigg(&s_{11}^{A_{2+n-\lambda}}+r_{(3+n-\lambda)1}^{1}+\cdots+r_{11}^{\lambda-1}+s_{21}^{A_{2}}+\Big(d-r_{(2+n-\lambda)1}^{1}\Big)\\
            &+\sum_{g=2}^{\lambda-1}(d-r_{21}^{g}),\cdots,s_{1(2m)}^{A_{2+n-\lambda}}+r_{(3+n-\lambda)2m}^{1}+\cdots+r_{1(2m)}^{\lambda-1}\\
            &+s_{2(2m)}^{A_{2}}+\Big(d-r_{(2+n-\lambda)2m}^{1}\Big)+\sum_{g=2}^{\lambda-1}\Big(d-r_{2(2m)}^{g}\Big)\Bigg),\\
            \\
            &\cdots,\\
            \\
            P_{n}=\Bigg(&s_{11}^{A_{n-\lambda}}+r_{(n+1-\lambda)1}^{1}+\cdots+r_{(n-1)1}^{\lambda-1}+s_{21}^{A_{n}}+\Big(d-r_{(n-\lambda)1}^{1}\Big)\\
            &+\sum_{g=2}^{\lambda-1}(d-r_{n1}^{g}),\cdots,s_{1(2m)}^{A_{n-\lambda}}+r_{(n+1-\lambda)2m}^{1}+\cdots+r_{(n-1)2m}^{\lambda-1}\\
            &+s_{2(2m)}^{A_{n}}+\Big(d-r_{(n-\lambda)2m}^{1}\Big)+\sum_{g=2}^{\lambda-1}\Big(d-r_{n(2m)}^{g}\Big)\Bigg)
        \end{split}
    \end{align}
    where $s_{1j}^{A_{i}}$ and $s_{2j}^{A_{i}}$ $(i=1,2,\cdots,n;\ j=1,2,\cdots,2m)$ present the classical results of particles $p_{1j}^{(i)}$ and $p_{2j}^{(i)}$ after $\left | 0  \right \rangle ,\cdots,\left | d-1  \right \rangle$ basis measurement. 
    \item All participants compute $P_{1}+P_{2}+\cdots+P_{n}= \left ( T_{1},\cdots,T_{2m} \right )$, where $\left ( T_{1}+T_{2},T_{3}+T_{4},\cdots,T_{2m-1}+T_{2m} \right )$ is the summation of the participants' inputs.
\end{enumerate}

\subsubsection{Correctness}\label{sec4_3_2}

Similar to the analysis in the Sect.\ref{sec4_2_2}, $A_{1},A_{2},\cdots,A_{n}$ $(n \ge 6)$ obtain the value of Eq.(\ref{eq_16}) and compute 
\begin{align}
\begin{split}
    P_{1}+\cdots+P_{n}=\Bigg(&s_{11}^{A_{1+n-\lambda}}+r_{(2+n-\lambda)1}^{1}+\cdots+r_{n1}^{\lambda-1}+s_{21}^{A_{1}}+\Big(d-r_{(1+n-\lambda)1}^{1}\Big)\\
    &+\sum_{g=2}^{\lambda-1}(d-r_{11}^{g})+\cdots+s_{11}^{A_{n-\lambda}}+r_{(n+1-\lambda)1}^{1}+\cdots+r_{(n-1)1}^{\lambda-1}\\
    &+s_{21}^{A_{n}}+\Big(d-r_{(n-\lambda)1}^{1}\Big)+\sum_{g=2}^{\lambda-1}(d-r_{n1}^{g}),\cdots,s_{1(2m)}^{A_{1+n-\lambda}}\\
    &+r_{(2+n-\lambda)2m}^{1}+\cdots+r_{n(2m)}^{\lambda-1}+s_{2(2m)}^{A_{1}}+\Big(d-r_{(1+n-\lambda)2m}^{1}\Big)\\
    &+\sum_{g=2}^{\lambda-1}\Big(d-r_{1(2m)}^{g}\Big)+\cdots+s_{1(2m)}^{A_{n-\lambda}}+r_{(n+1-\lambda)2m}^{1}+\cdots\\
    &+r_{(n-1)2m}^{\lambda-1}+s_{2(2m)}^{A_{n}}+\Big(d-r_{(n-\lambda)2m}^{1}\Big)+\sum_{g=2}^{\lambda-1}\Big(d-r_{n(2m)}^{g}\Big)\Bigg)\\
    =\Big(&x_{11}'+\cdots+x_{n1}',\cdots,x_{1(2m)}'+\cdots+x_{n(2m)}\Big),
\end{split}
\end{align}
where $x_{i(2j-1)}'+x_{i(2j)}'=x_{ij}$ $(i=1,2,\cdots,n;\ j=1,2,\cdots,m)$, and $\Big((x_{11}'+\cdots+x_{n1}')+(x_{12}'+\cdots+x_{n2})',\cdots,\big(x_{1(2m-1)}'+\cdots+x_{n(2m-1)}'\big)+\big(x_{1(2m)}'+\cdots+x_{n(2m)}'\big)\Big)$ is the summation of all participants' inputs.

\subsubsection{Security Analysis}\label{sec4_3_3}

The collusive attack from two participants referred in Sect.\ref{sec3} and the outside attack is invalid for the improved protocol, whose analysis is similar to Sect.\ref{sec4_2_3}. For the collusive attack from $n-2$ participants, except for $A_{k}$ and $A_{l}$ $(k,l=1,2,\cdots,n$ and $k < l)$, we also suppose the remaining $n-2$ parties as the dishonest participants. If $l=k+\lambda$ or $l=k+n-\lambda$, the remaining $n-2$ parties can get the value of $s_{1j}^{A_{k}}+s_{2j}^{A_{l}}$ in Step 5, but they can't obtain $x_{kj}$ and $x_{lj}$ $(j=1,2,\cdots,2m)$ since loss of the exact value of $s_{1j}^{A_{k}}$ and $s_{2j}^{A_{l}}$. If $l \ne k+\lambda$ and $l \ne k+n-\lambda$, the remaining parties can obtain the value of $s_{1j}^{A_{l}}+r_{kj}$ or $s_{2j}^{A_{l}}+r_{kj}$, but they also can't restore $x_{lj}$ since loss of the knowledge of $r_{kj}$. Consequently, any $n-2$ participants have no chance to reap the remaining two parties' secret inputs and the security of the improved protocol can be guaranteed.

\section{Discussion and Conclusion}\label{sec5}

In the improved protocol proposed in this paper, we have not only addressed the security vulnerabilities present in Wu's protocol, but also improved the efficiency. The comparison results on efficiency are showed as follows. Here, the qubit efficiency is defined beforehand as
\[\eta =\frac{c}{q+b},\]
where $c$ denotes the total number of the classical plaintext message bits, $q$ denotes the total number of qubits used in quantum scheme and $b$ denotes the number of classical bits exchanged for decoding the message. In the multi-party improved protocol, $4mn$ particles are used for encoding. Participants need to publish $(\lambda-1)nt$ classical bits in Step 4, and announce $2mn$ classical bits in Step 5. Accordingly, the qubit efficiency is $\eta =\frac{2m}{\big((\lambda-1)t+2m\big)n}$, when $t$ is small enough, excellent efficiency can be achieved.

In conclusion, in this paper, improved protocols based on $d$-dimensional Bell states is proposed, whose security is verified in detail, guaranteed that it can resist both outside attack and participant attack. In addition, the improved protocol removes the decoy photons technology to detect eavesdropper, replaced by the method that checks whether the particles are collapsed during the transmission in Step 4 of the improved four or five-party protocol and the improved multi-party protocol.

\bmhead{Acknowledgments}
This work was supported by the Guangdong Basic and Applied Basic Research Foundation (Grant Nos. 2021A1515011985, 2022A1515140116) and the National Natural Science Foundation of China (Grant No. 61902132).

\bibliography{sn-bibliography}
\end{document}